\documentclass[a4paper]{PoS}

\usepackage{lineno}

\title{Data Quality Monitoring system in the Baikal-GVD experiment}

\ShortTitle{DQM in the Baikal-GVD}

\author{A.D.~Avrorin$^{,a}$, A.V.~Avrorin$^a$, V.M.~Aynutdinov$^a$, R.~Bannash$^g$, I.A~Belolaptikov$^b$, V.B.~Brudanin$^b$, N.M.~Budnev$^c$, G.V.~Domogatsky$^a$, A.A.~Doroshenko$^a$, \speaker{R.~Dvornick\'y$^{b,h}$}, A.N.~Dyachok$^c$, Zh.-A.M.~Dzhilkibaev$^a$, L. Fajt$^{b,h,i}$, S.V~Fialkovsky$^e$, A.R.~Gafarov$^c$, K.V.~Golubkov$^a$, N.S.~Gorshkov$^b$, T.I.~Gress$^c$, R.~Ivanov$^b$, K.G.~Kebkal$^g$, O.G.~Kebkal$^g$, E.V.~Khramov$^b$ , M.M.~Kolbin$^b$, K.V.~Konischev$^b$, A.V.~Korobchenko$^b$, A.P.~Koshechkin$^a$, A.V.~Kozhin$^d$, M.V.~ Kruglov$^b$, M.K.~Kryukov$^a$, V.F.~Kulepov$^e$, M.B.~Milenin$^e$, R.A.~Mirgazov$^c$, V.~Nazari$^b$, \fbox{A.I.~Panfilov$^a$}, D.P.~Petukhov$^a$ E.N.~Pliskovsky$^b$, M.I.~Rozanov$^f$, E.V.~Rjabov$^c$, V.D.~ Rushay$^b$, G.B.~Safronov$^b$, B.A.~Shaybonov$^b$, M.D.~Shelepov$^a$, F.~\u{S}imkovic$^{b,h,i}$, A.V.~Skurikhin$^d$, A.G.~Solovjev$^b$, M.N.~ Sorokovikov$^b$, I.~\u{S}tekl$^i$, O.V.~Suvorova$^a$, E.O.~Sushenok$^b$, V.A.~Tabolenko$^c$, B.A.~Tarashansky$^c$, S.A.~Yakovlev$^g$\\
$^a$ Institute for Nuclear Research, Russian Academy of Sciences, Moscow, 117312 Russia\\
$^b$ Joint Institute for Nuclear Research, Dubna, 141980 Russia\\
$^c$ Irkutsk State University, Irkutsk, 664003 Russia\\
$^d$ Institute of Nuclear Physics, Moscow State University, Moscow, 119991 Russia\\
$^e$ Nizhni Novgorod State Technical University, Nizhni Novgorod, 603950 Russia\\
$^f$ St. Petersburg State Marine Technical University, St. Petersburg, 190008 Russia\\
$^g$ EvoLogics Gmbh, Germany\\ 
$^h$ Comenius University, Mlynska Dolina F1, Bratislava, 842 48 Slovakia\\
$^i$ Czech Technical University in Prague, Prague, 128 00 Czech Republic\\
E-mail: \email{dvornicky@dnp.fmph.uniba.sk}
}

\abstract{The quality of the incoming experimental data has a significant importance for both analysis and running the experiment. The main point of the Baikal-GVD DQM system is to monitor the status of the detector and obtained data on the run-by-run based analysis. It should be fast enough to be able to provide analysis results to detector shifter and for participation in the global multi-messaging system.}

\FullConference{36th International Cosmic Ray Conference -ICRC2019-\\
		July 24th - August 1st, 2019\\
		Madison, WI, U.S.A.}

\begin{document}

\section{Introduction}

There are several requirements to the monitoring system for the Baikal-GVD experiment~\cite{refId0}. It should be as simple as possible to be fast in running and clear enough for use after reading the manual. But at the same time the system should be reliable and provide monitoring of all crucial parameters of the detector. In general, parameters to be monitored could be divided into two main groups. The first one monitors the statistical parameters of the detector to register muons. Since the trigger of the detector is extremely basic and firing if any neighbor pair of channels (or OM - optical module) generates signals exceeding at least high ($\sim$5 p.e.) and low ($\sim$2 p.e.) charge thresholds, the main source of events passing this trigger are atmospheric muons and random noise~\cite{Avrorin2014}. The second set of parameters is charge related. It analyses charge distributions of both trigger and non-trigger signals recorded by a given channel. The minimum $\chi^2$ fit algorithm is applied while performing the fit of the distributions to a corresponding expected function. In addition to the parameters mentioned above the trigger efficiency of the cluster is also monitored.

\section{Statistical parameters}

There are three parameters monitored within this set: exponential, uniformity and Poisson. For each of them the monitoring system analyses shapes of distributions for all events in the run under consideration. The shape analysis includes the fit of the distribution with the expected function following by the estimation of the fit quality. These parameters are analysed for each channel, section, and the whole cluster separately~\cite{Avrorin2014}. Besides the fit quality estimation using the $\chi^2/NDF$ parameter, the amount of bins that deviate from the obtained fit-function is calculated. Only statistical uncertainty is considered to estimate the number of standard deviations for the given bin of the distribution.

\subsection{Exponential distribution}

\begin{figure}
\centering
\includegraphics[width=.49\textwidth]{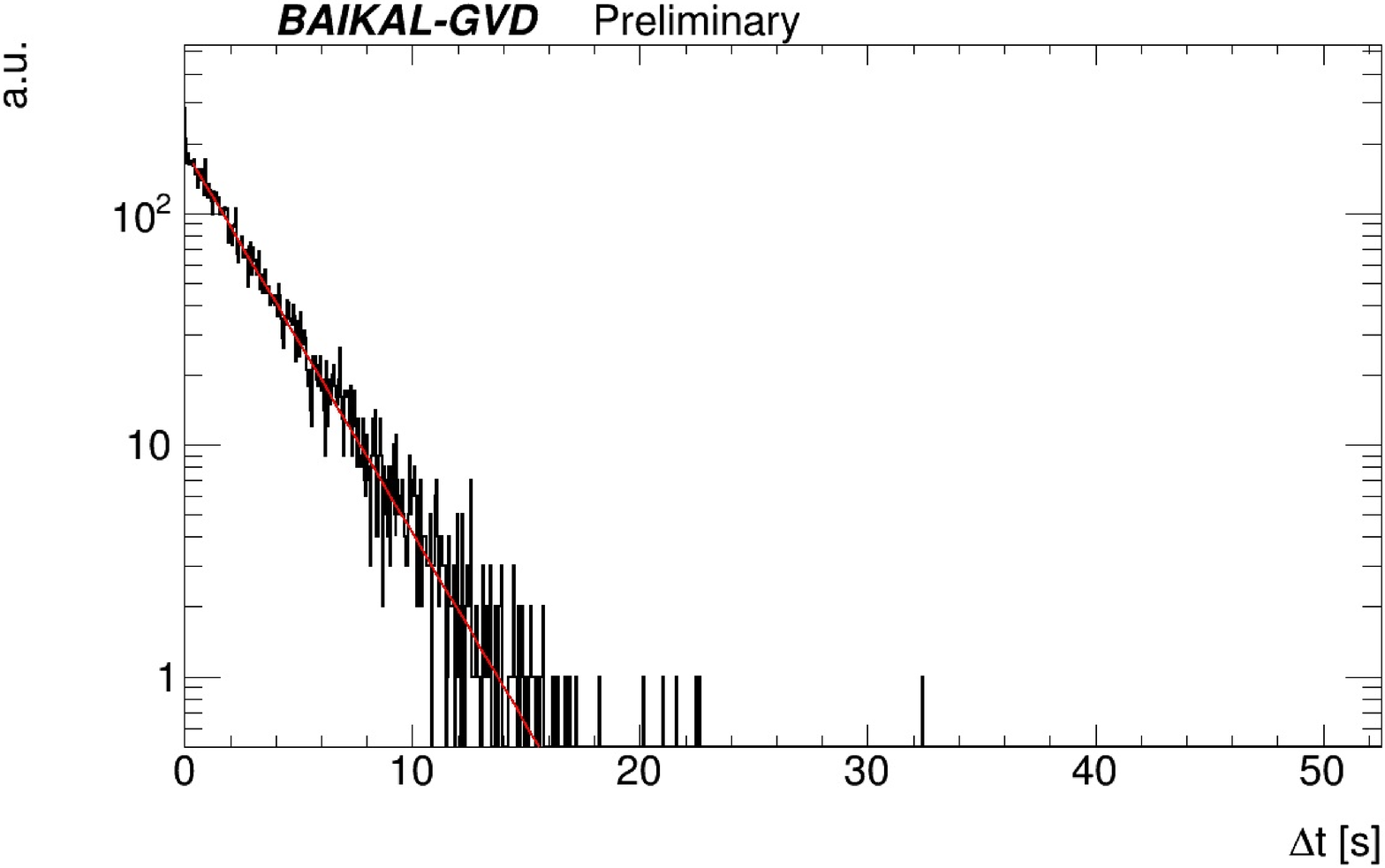}
\includegraphics[width=.49\textwidth]{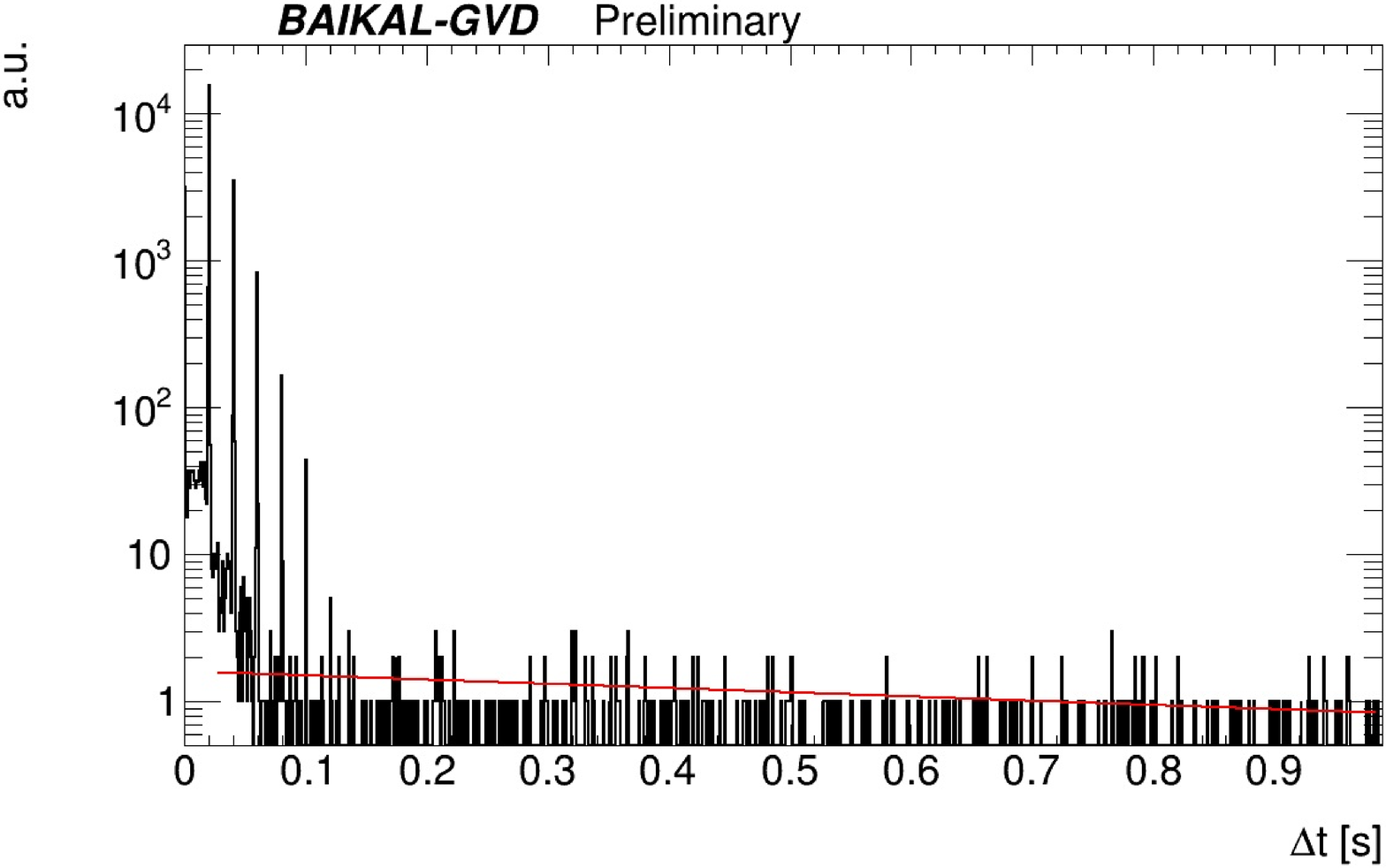}
\caption{Exponential distributions for atmospheric muons and random noise for some selected channels.
Right: for calibration run with LED matrix.}
\label{fig:exp}
\end{figure}

Since the main type of events passing the trigger are atmospheric muons and random noise the time difference between two neighbor events should be described by the exponential function (see Fig.~\ref{fig:exp} left). The bin width is chosen to suppress possible bin-to-bin statistical fluctuations. Due to the dead time of the trigger operation the maximum of the distribution is at $\sim$1 ms and events with a time difference lower than 1 ms corresponds to statistics recorded by the second event buffer of the trigger system. Therefore the next bin to the maximum one is chosen as the low edge of the fit range and till the last bin. The calibration systems used for PMT calibration such as LED matrices or laser sources due to the fixed operation frequency values should deteriorate the exponential shape of the atmospheric muons and random noise (see Fig.~\ref{fig:exp} right) and therefore decrease the fit quality and increase the number of bins with very high deviation value from the obtained with the fit-function.

\subsection{Uniformity distribution}

\begin{figure}
\centering
\includegraphics[width=.49\textwidth]{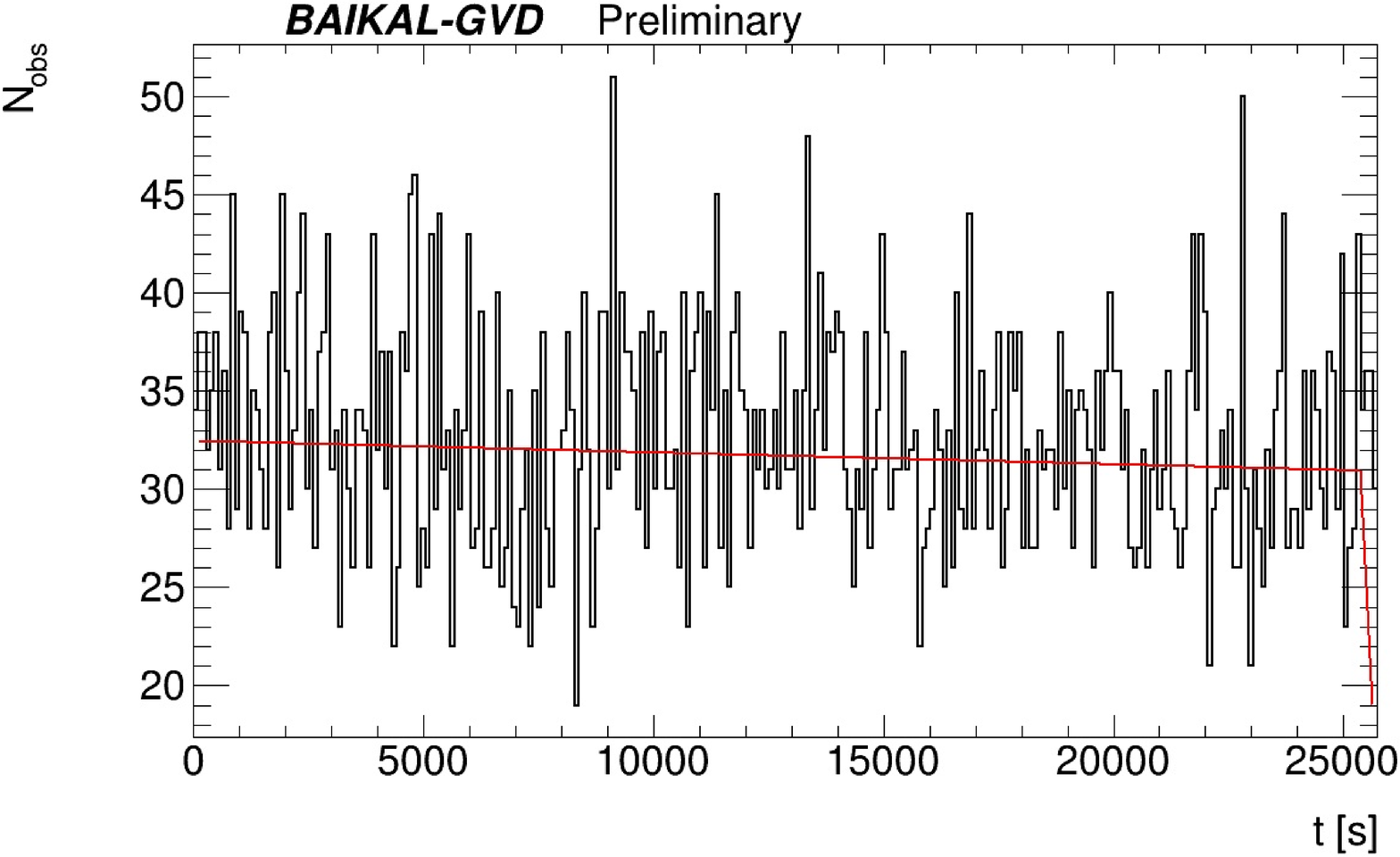}
\includegraphics[width=.49\textwidth]{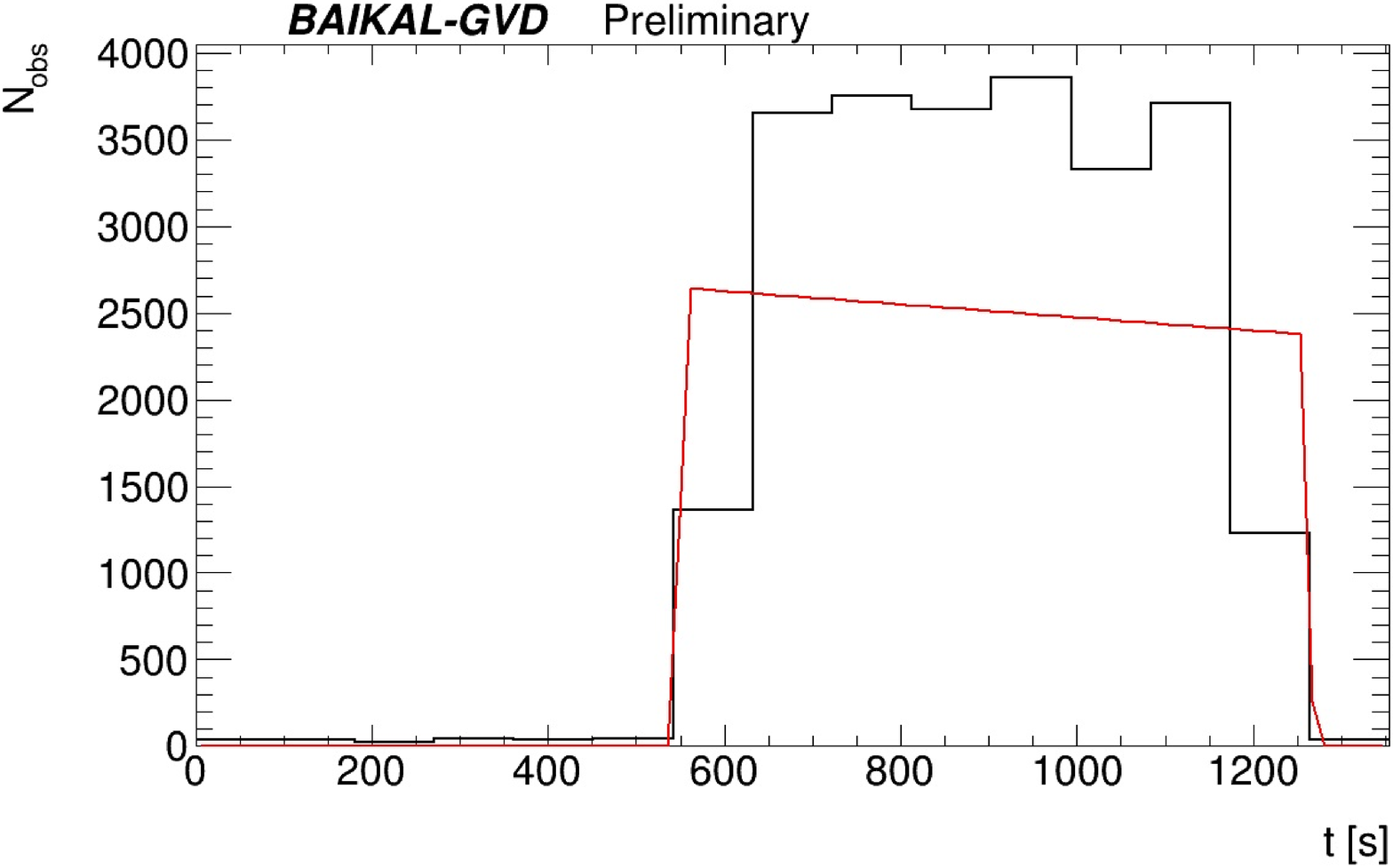}
\caption{Uniformity distributions for atmospheric muons and random noise for some selected channels.
Right: the same for calibration run with LED matrix.}
\label{fig:uni}
\end{figure}

It is expected that the distribution of the rate of atmospheric muons and random noise is linear with possible slope due to the environment (temperature, season, water transparency etc) or detector effects (current fluctuations etc.) and should be described by the first order polynomial function. The expected cluster rate is $∼\sim$50-200 events/sec. As well as for the exponential distribution the bin width for the corresponding uniformity distribution is chosen to suppress possible bin-to-bin statistical fluctuations (see Fig.~\ref{fig:uni} left). As it is seen in Fig.~\ref{fig:uni} (right) events with LED calibration matrices applied show a significant increase in rate with very clear boundaries thus the fit quality decreases.

\subsection{Poisson distribution}

\begin{figure}
\centering
\includegraphics[width=.49\textwidth]{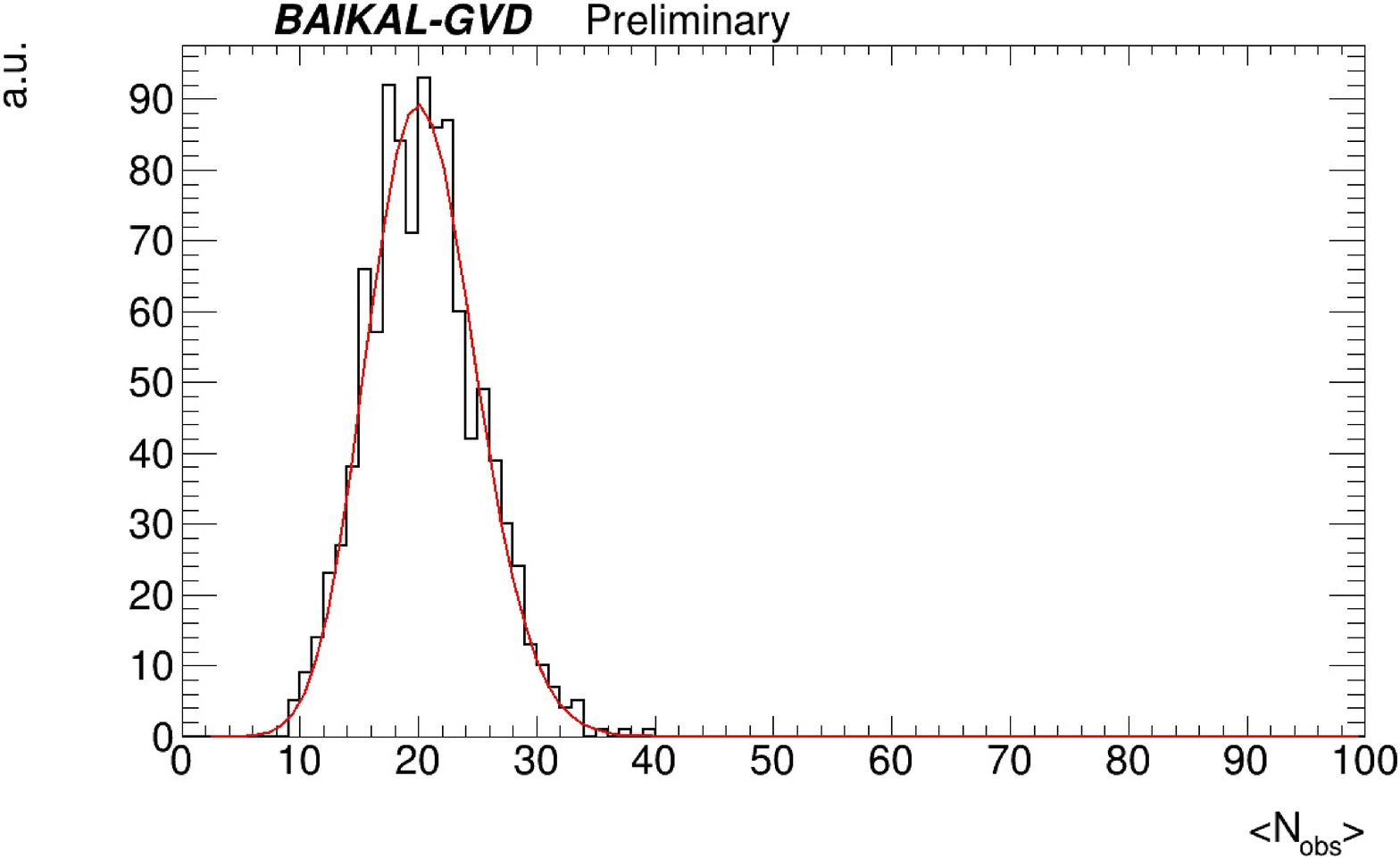}
\includegraphics[width=.49\textwidth]{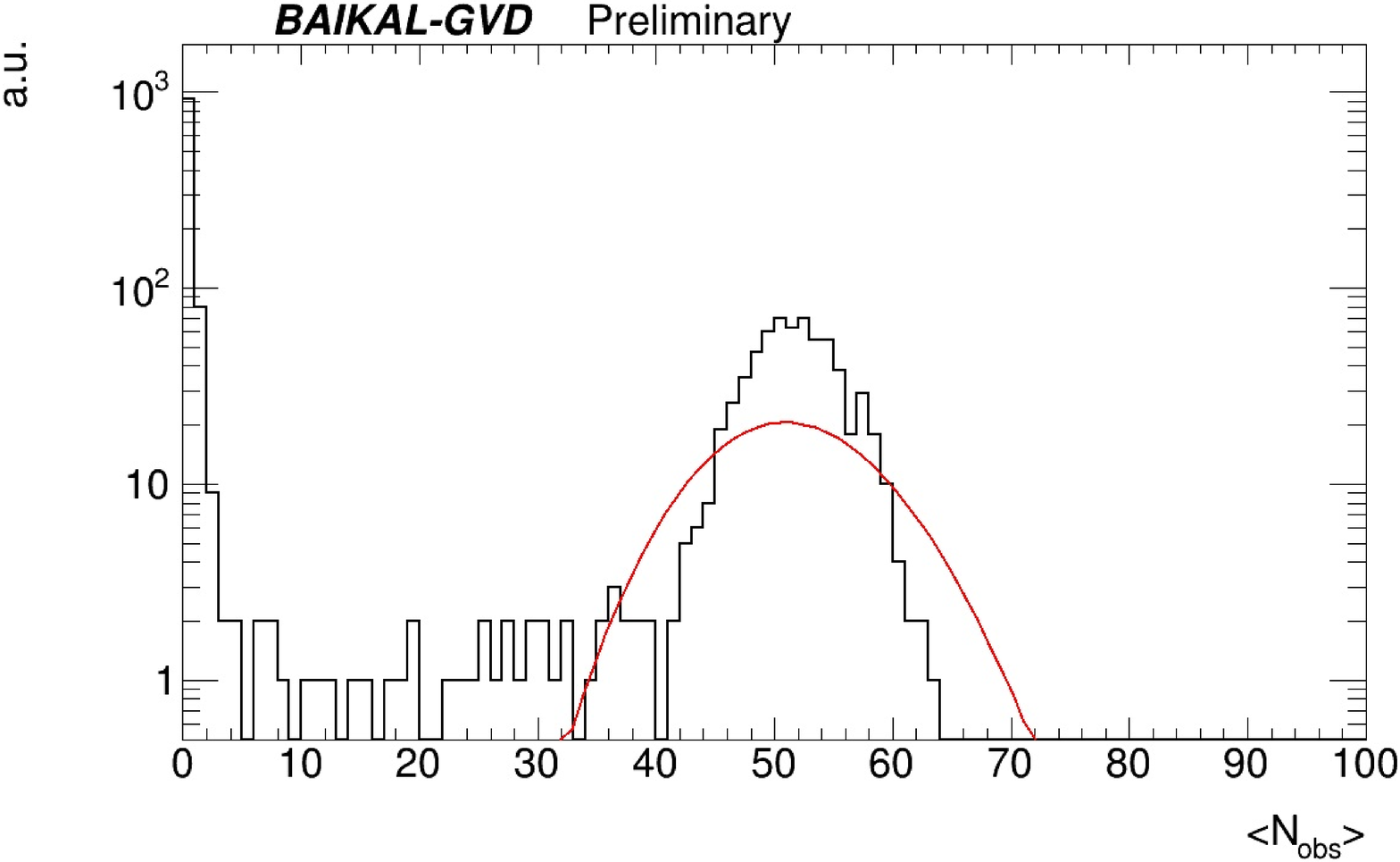}
\caption{Rate distributions for atmospheric muons and random noise for some selected channels fitted with a Poisson function. 
Right: the same for calibration run with LED matrix.}
\label{fig:poi}
\end{figure}

The expected value of the event rate for the given cluster size ($\sim$50-200 events/sec per cluster) should follow Poissonian distribution for any fixed time interval. Thus, for each run the time interval is chosen to have $\sim$20 recorded events on average and the distribution of the number of events per this time range is tested for the poisson statistics (see Fig.~\ref{fig:poi} left). There is clear effect of the calibration run with the LED matrix when the average number of events per given time interval differs from poissonian shape and the mean value is lower than expected number ($\sim$20) of recorded events (see Fig.~\ref{fig:poi} right).

\section{Charge distribution}

Another important set of parameters for the run quality monitoring is referred to the distributions of the charge from the PMT. Therefore the data quality is estimated for given channel only and considered in summarizing algorithm. The low threshold for the charge is the so-called filter threshold that is about half of the single photoelectron value. It is possible to analyse separately channels that produced trigger in the given event from all other ones as well as full combination of signals from channel.

\subsection{Distribution for non-trigger signals}

\begin{figure}
\centering
\includegraphics[width=.49\textwidth]{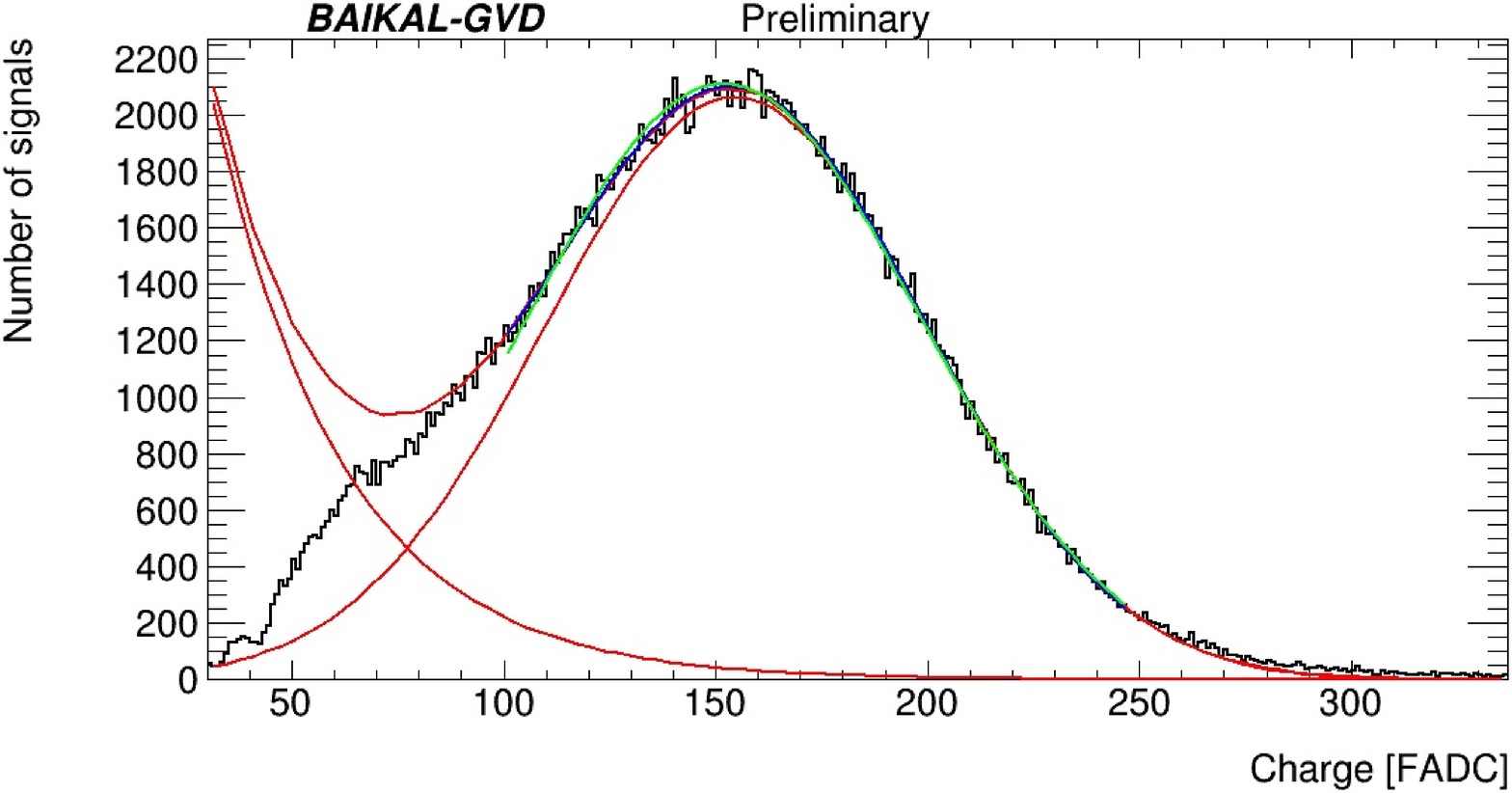}
\includegraphics[width=.49\textwidth]{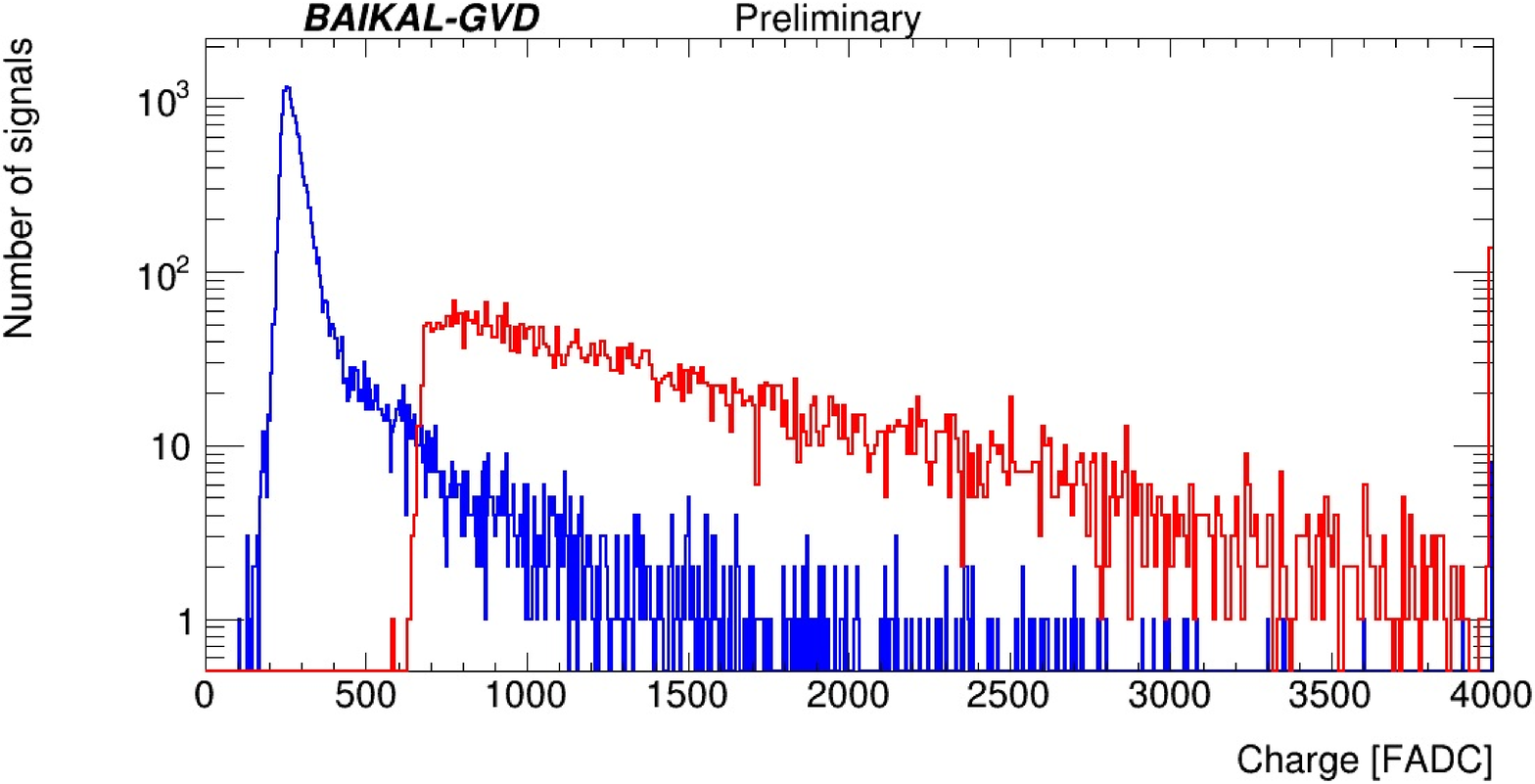}
\caption{Left: the charge distribution for some selected non-trigger channel; Right: charge values for some selected channel participating in the trigger.}
\label{fig:1pe}
\end{figure}

Any charge from the PMT exceeding the so-called "filter threshold" ($\sim$0.5 p.e.) is recorded. Selecting only channels that did not fire the trigger for a given event we can determine the position of the 1 p.e. peak for given channel in the current run produced mainly by chemi luminescence. As one can see in Fig.~\ref{fig:1pe} (left) along with very clear 1 p.e. gaussian distribution with the mean value at around 160 FADC, there is some effect from dark currents of the PMT on the left side of the 1 p.e. distribution. Also there is negligible impact from the 2 p.e. distribution that is seen on the right side of the 1 p.e. one. To describe the combined distribution the sum of two gaussian functions for the 1 and 2 p.e. distributions and an exponential function for the dark current part was used.

\begin{figure}
\centering
\includegraphics[width=.7\textwidth]{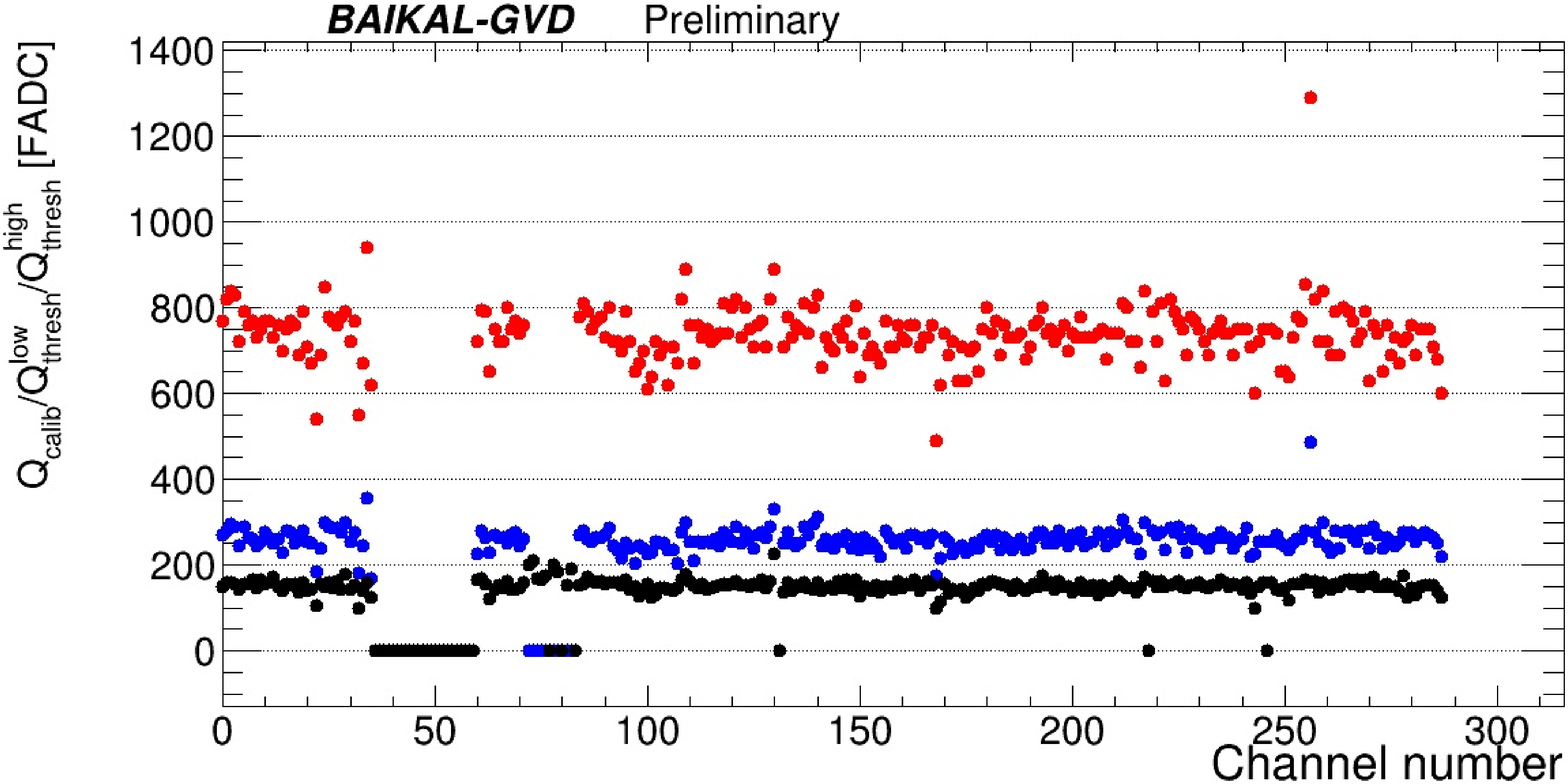}\\
\includegraphics[width=.7\textwidth]{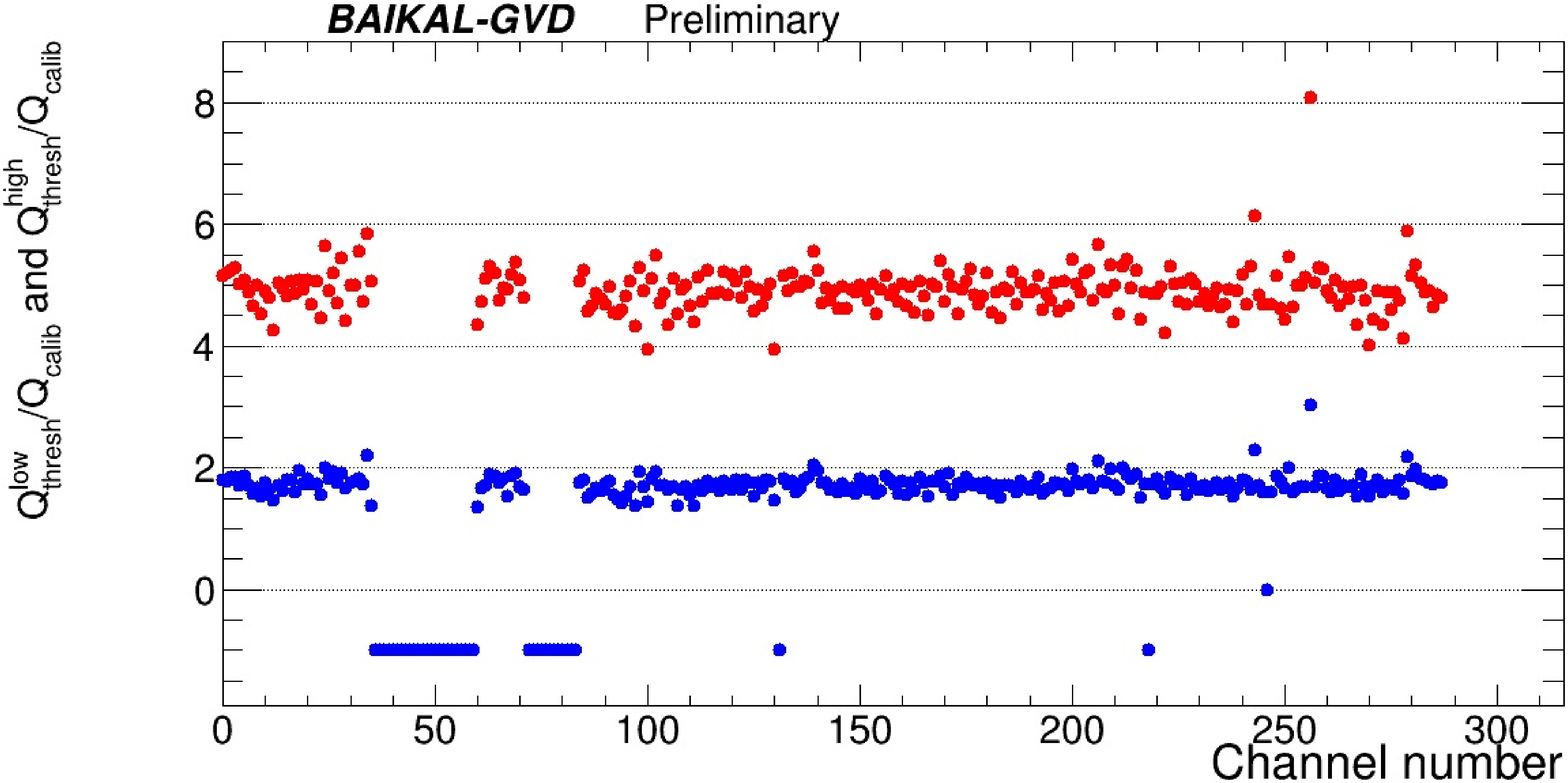}
\caption{Top: charge calibration (black), trigger low (blue) and high (red) threshold values for channels in some selected run; Bottom: Ratios of trigger low (blue) and high (red) threshold to charge calibration values for channels in some selected run.}
\label{fig:qcalibrun}
\end{figure}

\begin{figure}
\centering
\includegraphics[width=.7\textwidth]{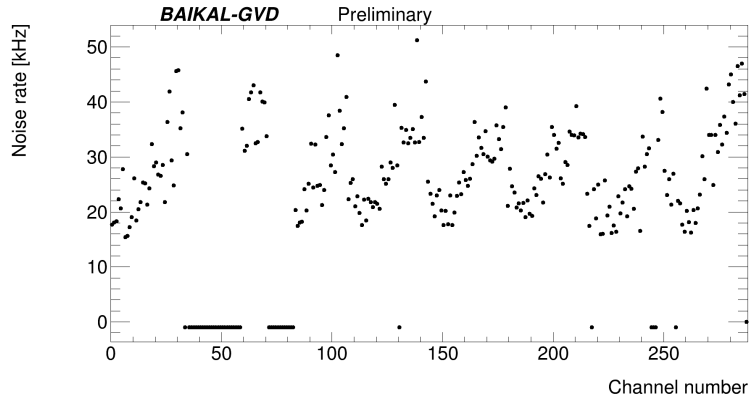}
\caption{The noise rates for channels in some selected run.}
\label{fig:noiserun}
\end{figure}

Combined information for all channels in run is used for control of PMTs work stability. As one can see in Fig.~\ref{fig:qcalibrun} (top) the 1 p.e. calibration values (black points) are rather stable over all channels in the run. The spread of the values is due to the manual adjustment of the mean value for each channel and the difference of the PMTs themselves. In addition to calibration behavior it is possible estimate the stability of the noise rate of the channel. Statistics recorded by given channel is proportional to the noise rate of this channel. This noise rate is calculated as the distribution integral from 0.5 p.e. for each channel (see Fig.~\ref{fig:noiserun}). It is expected that this values should increase towards the higher vertical position of the channel on the string~\cite{RastVLVNT18}. The noise rate for channel is closely correlated to the calibration value and thus it is expected that at every depth should be stable from channel to channel. The deviation of noise related value for each channel from the average over the given depth level is the parameter for the channel quality estimation.

\subsection{Distribution for trigger signals}

The values of low and high trigger threshold are set to $\sim$2 and $\sim$5 photoelectrons. It is very important to make sure that these values are stable across the run and season. In Fig.~\ref{fig:1pe} (right) it is seen that the shapes of the corresponding distributions have thresholds that are extracted and used for monitoring (see Fig.~\ref{fig:qcalibrun} (top) low (blues) and high (red) thresholds). Since the exact value of the threshold in p.e. depends on the charge calibration of the PMT, the ratio of the threshold value in the given channel to its charge calibration mean value are monitored. It is expected that these ratios are stable across all channels within both run and season (see Fig.~\ref{fig:qcalibrun} bottom) and some deviations are possible due to the same reasons as for the calibrations itself.



\subsection{Shape comparison for all signals}

\begin{figure}
\centering
\includegraphics[width=.49\textwidth]{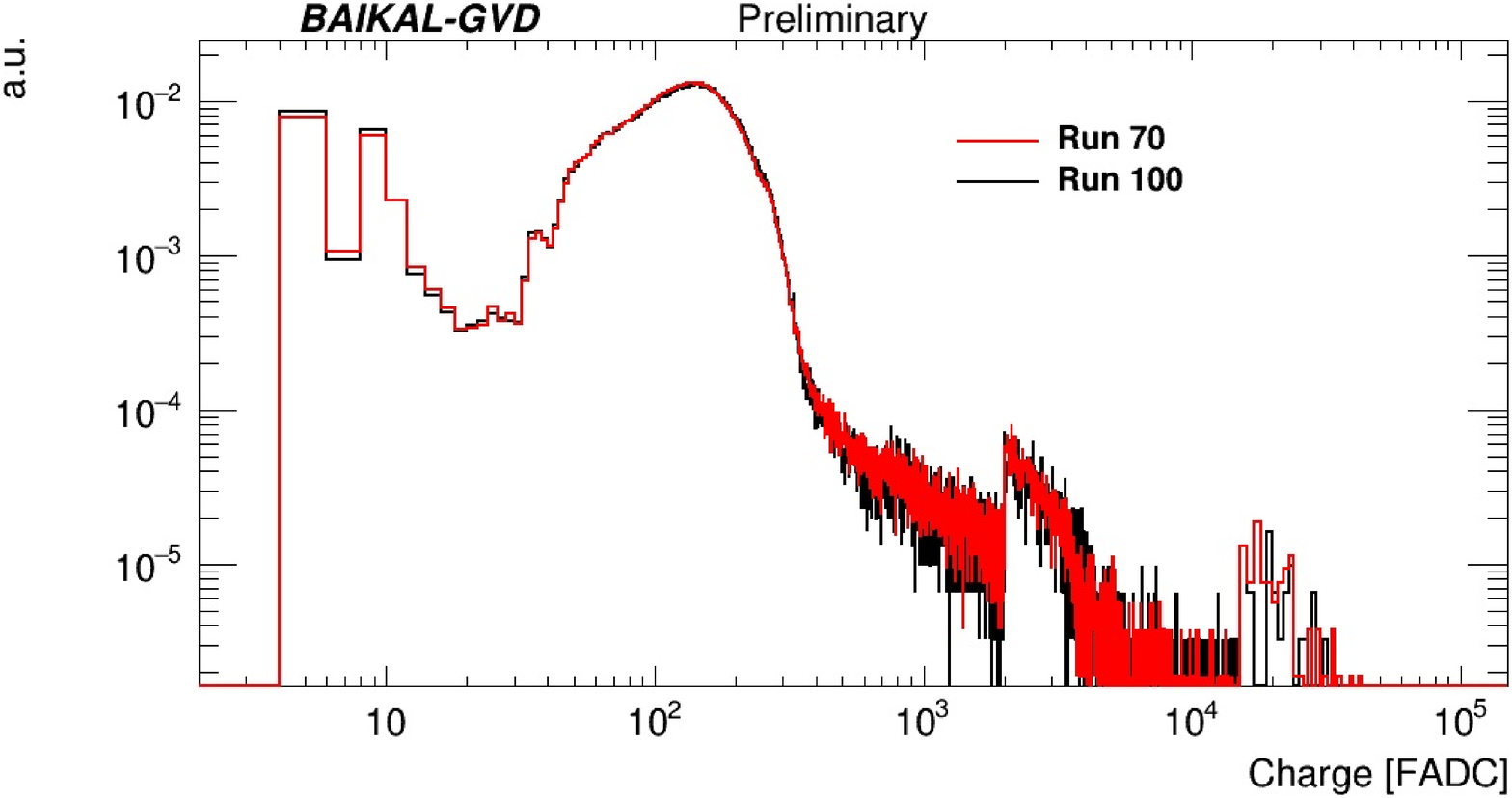}
\includegraphics[width=.49\textwidth]{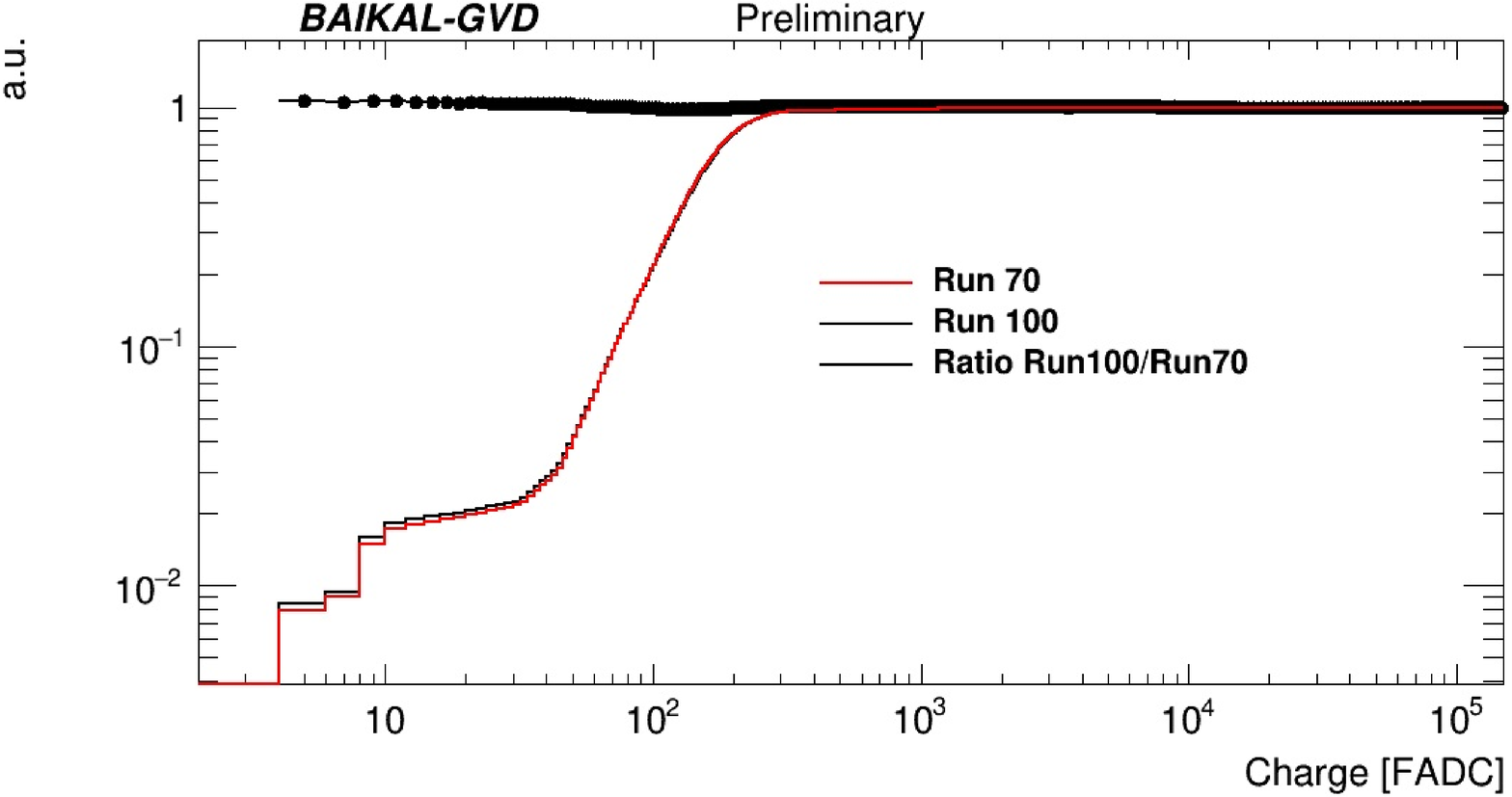}
\caption{Charge distributions (left) and cumulative distributions obtained from them (right) for basic (red) and considered (blue) runs correspondingly. Ratio of charge distributions is shown in black on right panel.}
\label{fig:shapes}
\end{figure}

In addition to calibration and threshold values the general comparison of the charge distribution shapes are performed. Shapes of distributions of any charge deposited in the channel from the filter threshold up to 1000 p.e. are compared. Distribution from the run under analysis is compared to a so-called basic run that is well known to have good channel performance. The shapes are compared via cumulative distributions normalized to 1. The relative impact of the integral in a given range starting from 0 is the estimation parameter. Fig.~\ref{fig:shapes} shows the example of charge distributions (left) and cumulative distributions obtained from them (right) for basic (red) and considered (blue) runs correspondingly. Number of events with charge value deposited in channel more than 100 p.e. should not exceed 100 events, otherwise the channel is masked as flashed by LED matrix.

\section{Gantt-plots}

\begin{figure}
\centering
\includegraphics[width=.7\textwidth]{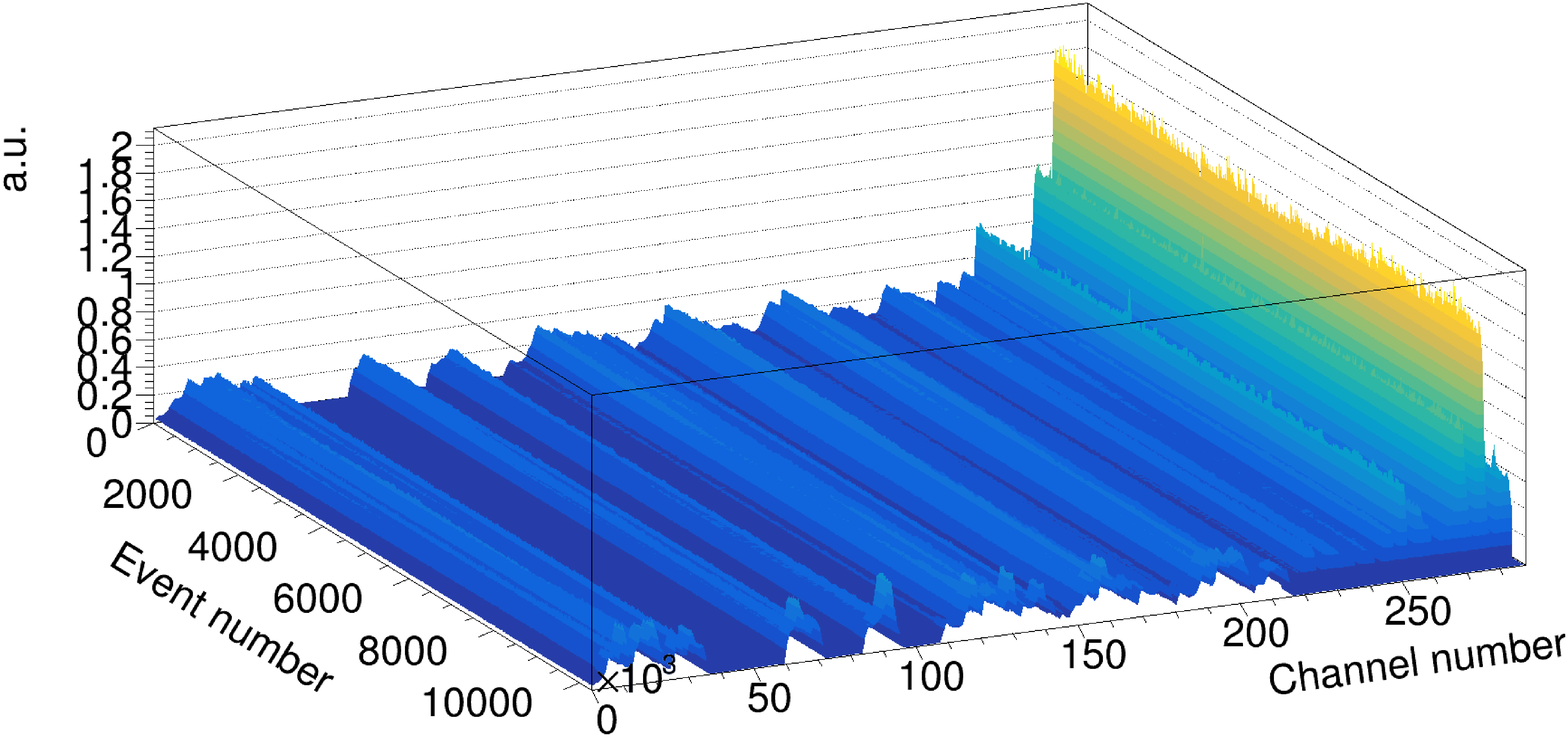}\\
\includegraphics[width=.7\textwidth]{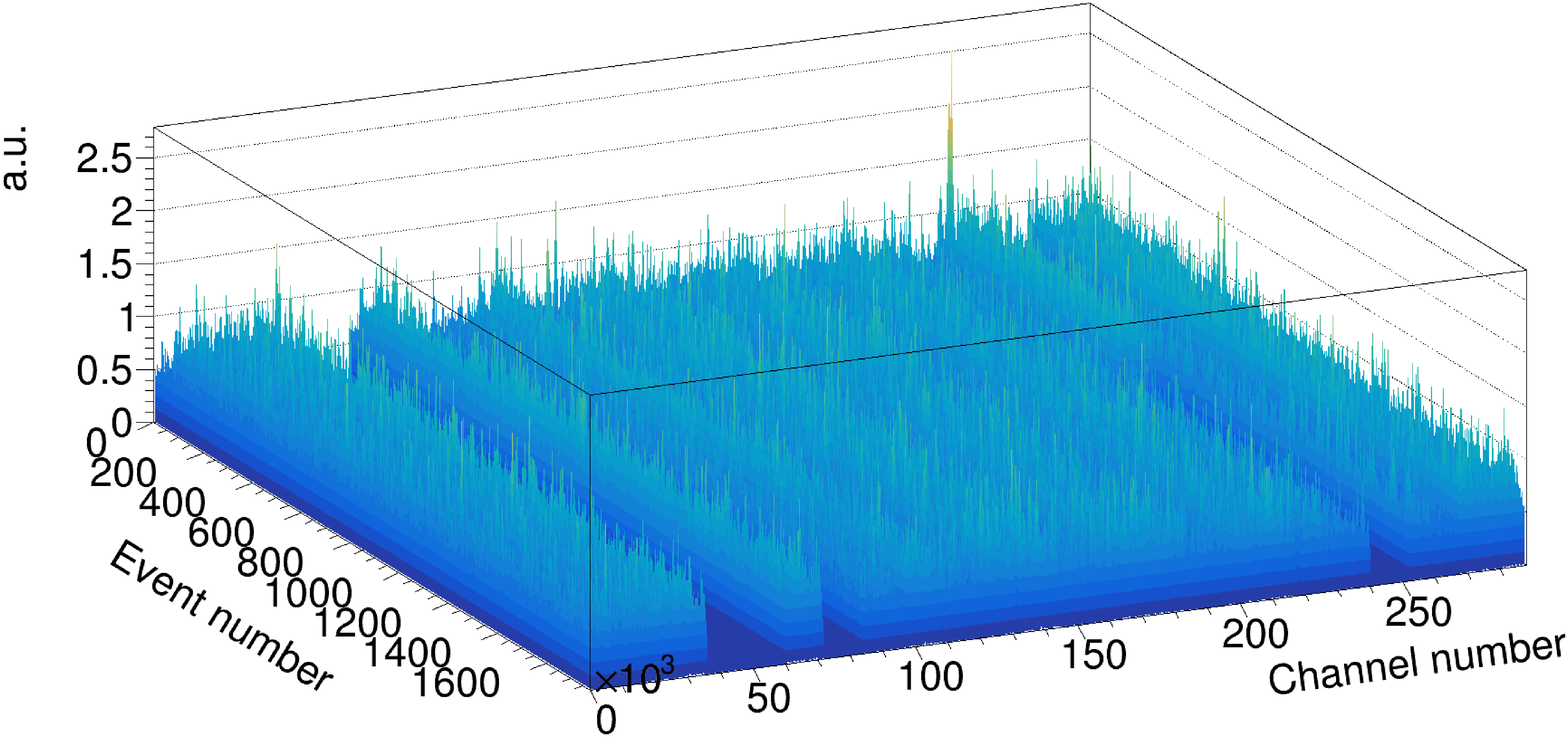}
\caption{Top: the Gantt-plot used for selection of events with the LED matrix switched on in each channel.
Bottom: the Gantt-plot used for selection events with stable event rate in each channel.}
\label{fig:gantt}
\end{figure}

To analyse as much events as possible with the given cluster configuration the Gantt-plots are used for two main purposes. The first one is to avoid events with the LED matrix switched on. In this case the number of events recorded by cluster increases by several orders of magnitude thus to save the time processing the number events range tested for this issue is 10\,000. The Gantt-plott normalized to number of events is seen in Fig.~\ref{fig:gantt} (top) that channels with LED matrix switched on exceed 0.5 number of entries per bin. Thus in standard conditions we expect maximum a doubling of the rate during the run.

For the second purpose the other Gantt-plot was used. This plot contains 100 thousand bins for the axis corresponding to event number and is also normalized to the total number of events in the run. Such high granularity allows to select events range with the channel working properly. Assuming that each channel should not have more than one signal higher than zero the upper limit of the bin content is set to 1 (see Fig.~\ref{fig:gantt} bottom).

\section{Quality estimation algorithm}

After estimation of the run using parameters described above the solution algorithm is applied. The main point of this algorithm is to provide for the analysis procedure information about quality status at each level of the detector from channel to the whole cluster. There are eight ranking markers for data quality status (codes):

\begin{description}
 \item[0] -- excluded by configuration (defined by detector performance experts)
 \item[1] -- empty data (<100 events in channel)
 \item[2] -- good (successfully passed criteria)
 \item[3] -- normal (passed criteria with some warnings)
 \item[4] -- bad (failed criteria)
 \item[5] -- good data but LED is detected
 \item[6] -- normal data but LED is detected
 \item[7] -- bad data but LED is detected
\end{description}

The decision chain is the following: channel $\rightarrow$ section $\rightarrow$ string $\rightarrow$ cluster levels. At any level (channel, section and cluster) for exponent, uniformity, Poisson and Charge (channel level only):

\begin{enumerate}
 \item Fit quality is estimated via $\chi^2/NDF$ with threshold values for good (<2), normal (<4) and bad (>4) data
 
 \item Deviation of each bin value from fit-function using following condition with threshold values $R_1$= 25\% and $R_2$=1\% for good data and $R_1$= 50\% and $R_2$=5\% for normal data correspondingly (any other case $\rightarrow$ bad data): $N_{Normal}^{Bins}/N_{Total}^{Bins}<R_1$ and $N_{Bad}^{Bins}/N_{Total}^{Bins}<R_2$
 
 \item The deviation of the noise rate for each channel from the average over the given depth level: <3$\sigma_{noise}^{level}$) for good data, <5$\sigma_{noise}^{level}$ for normal data and bad (>5$\sigma_{noise}^{level}$) data
 
 \item The relative impact of the integral in the [0-x] range of the distribution for non-trigger signals: <15\% for good data, <25\% for normal data and >25\% for bad data data
 
\end{enumerate}

\section{Conclusions}

The data quality monitoring system for the Baikal-GVD experiment was developed. Using some examples of the data recorded during the 2016 season it was shown that using various parameters for data quality analysis allows to estimate the quality of obtained data from the level of the optical module and up to the whole cluster very efficiently.

\section{Acknowledgements}

This work was supported by the Russian Foundation for Basic Research (Grants 16-29-13032, 17-0201237).

\end{document}